\documentclass[aps,amssymb,twocolumn]{revtex4}
\usepackage{amssymb}
\usepackage{graphicx}
\usepackage{color}
\usepackage{textcomp}
\usepackage{ulem}
\begin{document}

\title{LQCD constrained magnetic field dependent coupling constant in an effective model}
\author{Shijun Mao}%
 \email{maoshijun@mail.xjtu.edu.cn}
\affiliation{School of Physics, Xi'an Jiaotong University, Xi'an, Shaanxi 710049, China}

\begin{abstract}
A magnetic field dependent coupling constant $G(eB)$ is investigated in the two-flavor magnetized NJL model. Based on LQCD results of the neutral (charged) pion mass spectra at vanishing temperature and finite magnetic field, we determine the $G(eB)=G^0(eB)$ ($G(eB)=G^+(eB)$) in the NJL model. $G^0(eB)$ and $G^+(eB)$ are both non-monotonic functions of magnetic fields, but they are different from each other. Furthermore, we calculate the pseudo-critical temperatures $T_{pc}(eB)$ of chiral restoration phase transition with $G^0(eB)$ and $G^+(eB)$ in the magnetized NJL model, respectively. The resulting $T_{pc}(eB)$ are non-monotonic functions of magnetic fields. In previous work, $G(eB)$ in the NJL model fitted from the chiral condensate or pseudo-critical temperature of LQCD simulations is a decreasing function of magnetic field. It can not explain the saturation behavior of mass spectra of neutral pion and decreasing behavior of mass spectra of charged pion with strong magnetic field. We conclude that a magnetic field dependent coupling constant $G(eB)$ in the NJL model can not simultaneously explain the reduction of pseudo-critical temperature of chiral restoration phase transition and the light meson mass spectra under external magnetic field.
\end{abstract}

\date{\today}
\maketitle

\section{Introduction}
	
Chiral symmetry is one of the most important symmetries in QCD, and it is related to the light hadron properties. In recent years, the study of magnetic field effects on the chiral restoration phase transition and light hadron properties attract much attention due to their close relation to the heavy ion collision and compact stars~\cite{lattice1t0,lattice2t0,lattice4t0,lattice5t0,lattice6t0,lattice7t0,lattice9t0,lattice1,lattice11,lattice2,lattice3,lattice4,lattice5,lattice6,lattice7,fukushima,hadron2,bf1,bf12,bf13,mao,bf2,bf3,bf4,bf5,bf51,bf52,bf8,bf9,bf11,db1,db2,db3,db5,db6,pnjl1,pnjl2,pnjl3,pnjl4,pqm,ferr1,ferr2,huangamm2,meijie,imcreview,ammmao,geb1,geb3,geb4,c1,c3,hadron1,qm1,qm2,sigma1,sigma2,sigma3,sigma4,l11,l2,l3,l33,l4,lqcd55, lattice3m0,lattice5m0,lqcd6,ding200800493,ding2022,njl2,meson,mfir,ritus5,ritus6,mao1,mao11,mao2,wang,coppola,phi,liuhao3,he,maocharge,maopion,yulang201005716,q1,q2,q3,q4,huangamm1,q5,q66,q7,q8,q9,q10,Chernodub2010qx,Chernodub2011mc,Callebaut2011uc,Ammon2011je,Cai2013pda,Frasca2013kka,Andreichikov2013zba,Liu2014uwa,Liu2015pna,Liu2016vuw,Kawaguchi2015gpt,Ghosh2016evc,Ghosh2017rjo,Luschevskaya2014mna,Luschevskaya2015bea,Ghosh,su3meson1,su3meson3,su3meson4,su3meson5,limao2023,cao}.
The LQCD simulations~\cite{lattice1t0,lattice2t0,lattice4t0,lattice5t0,lattice6t0,lattice7t0,lattice9t0,lattice1,lattice11,lattice2,lattice3,lattice4,lattice5,lattice6,lattice7,ding2022} observe the inverse magnetic catalysis (IMC) phenomena, that is the chiral condensates of u and d quarks near the pseudo-critical temperature $T_{pc}$ of chiral restoration phase transition drop down with increasing magnetic field, and report the reduction of the pseudo-critical temperature $T_{pc}$ under external magnetic field. On the other side, LQCD results for neutral and charged light mesons ($\pi, K,\ \eta,\ \rho$) are available at zero temperature and finite magnetic field~\cite{l11,l2,l3,l33,l4,lqcd55,lattice3m0,lattice5m0,lqcd6,ding200800493,ding2022}. For instance, the masses of $\pi^0,\ K^0$ mesons decrease with magnetic fields, and they become saturated at strong magnetic field. The masses of $\pi^\pm,\ K^\pm$ mesons firstly increase and then decrease with magnetic fields. Note that $\pi^0,\ K^0$ mesons are the pseudo-Goldstone bosons corresponding to the spontaneous breaking of chiral symmetry, and the splitting between neutral and charged $\pi,\ K$ mesons represents the breaking of isospin symmetry under external magnetic field. On analytical sides, the physical explanation for LQCD results are not clear.

Without external magnetic field, the Nambu-Jona-Lasinio (NJL) model gives reasonable results for the phase transitions corresponding to chiral and isospin symmetries and the properties of related light mesons~\cite{njl1,njl2,njl3,njl4,njl5,zhuang,he1,he2,he3}. However, including external magnetic field, the NJL model meets some difficulty when comparing with LQCD results, such as in explaining the inverse magnetic catalysis phenomena, the reduction of pseudo-critical temperature of chiral restoration phase transition, and mass spectra of light mesons. Considering that the coupling constant $G$ plays very import role in determining the chiral restoration phase transition and mass spectra of light mesons, a running (magnetic field dependent) coupling constant $G(eB)$ has been introduced in the NJL model~\cite{geb1,meson,geb3,geb4,bf8,bf9,su3meson4,mao2,limao2023}. People fit $G(eB)$ by considering the inverse magnetic catalysis phenomena of chiral condensates~\cite{geb1,meson,geb3,geb4}, and $G(eB)$ is also determined by the decreasing pseudo-critical temperature of chiral restoration phase transition under external magnetic field~\cite{bf8,bf9,geb1,su3meson4,mao2,limao2023}. The two approaches give the decreasing $G(eB)$ with magnetic fields, which is consistent with each other. With such $G(eB)$, the decreasing (increasing) behavior of neutral (charged) pion mass spectra in weak magnetic field can be obtained, but the saturation behavior of $\pi^0$ meson mass spectra and decreasing behavior of $\pi^+$ meson mass spectra under strong magnetic field can not be realized~\cite{su3meson4,limao2023}. On the other side, a non-monotonic behavior of $G(eB)$~\cite{cao} is reported when considering the reduction of scaled $\pi^0$ meson mass $m_{\pi^0}(eB)/m_{\pi^0}(eB=0)$ with finite magnetic field and vanishing temperature~\cite{ding200800493}. The properties of chiral restoration phase transition and $\pi^+$ meson mass are not yet studied. It should be mentioned that the pion mass in vacuum used in Ref\cite{cao} is not the same as LQCD simulations~\cite{ding200800493}.

In this paper, to get a comprehensive understanding of the effect of magnetic field dependent coupling constant $G(eB)$ in the NJL model, we determine $G(eB)=G^0(eB)\ (G(eB)=G^+(eB))$ through the neutral (charged) pion mass spectra from LQCD simulations~\cite{ding200800493}, and calculate the pseudo-critical temperature of chiral restoration phase transition by using obtained $G(eB)$. We conclude that a magnetic field dependent coupling constant $G(eB)$ in the NJL model can not simultaneously explain both the reduction of pseudo-critical temperature of chiral restoration phase transition and the mass spectra of light mesons under external magnetic field.

The rest paper is organized as follows. Sec.\ref{sec:f} introduces our theoretical framework for chiral restoration phase transition and pion mass spectra in a magnetized NJL model. The numerical results and discussions are presented in Sec.\ref{sec:r}. Finally, we give the summary in Sec.\ref{sec:s}.

\section{Framework}
\label{sec:f}
The two-flavor NJL model is defined through the Lagrangian density in terms of quark fields $\psi$~\cite{njl1,njl2,njl3,njl4,njl5}
\begin{equation}
\label{njl}
{\cal L} = \bar{\psi}\left(i\gamma_\nu D^\nu-m_0\right)\psi+G \left[\left(\bar\psi\psi\right)^2+\left(\bar\psi i\gamma_5{\vec \tau}\psi\right)^2\right].
\end{equation}
Here the covariant derivative $D_\nu=\partial_\nu+iQ A_\nu$ couples quarks with electric charge $Q=diag (Q_u,Q_d)=diag (2e/3,-e/3)$ to the external magnetic field ${\bf B}=(0, 0, B)$ in $z$-direction through the potential $A_\nu=(0,0,Bx_1,0)$. $m_0$ is the current quark mass, which determines the explicit breaking of chiral symmetry. $G$ is the coupling constant in scalar and pseudo-scalar channels, which determines the spontaneous breaking and restoration of chiral symmetry and isospin symmetry.

The order parameter of chiral restoration phase transition is the chiral condensate $\langle{\bar{\psi}} \psi \rangle $ or (dynamical) quark mass $m_q=m_0-2G\langle{\bar{\psi}} \psi \rangle $. It is determined by the gap equation,
\begin{eqnarray}
1-2GJ_1(m_q)&=&\frac{m_0}{m_q},
\label{gap}
\end{eqnarray}
and
\begin{eqnarray}
J_1(m_q) &=& 3\sum_{f,n}\alpha_n \frac{|Q_f B|}{2\pi} \int \frac{d p_3}{2\pi} \frac{1-2F(E_f)}{ E_f},
\end{eqnarray}
with the summation over all quark flavors $f$ and quark Landau energy levels $n$, spin factor $\alpha_n=2-\delta_{n0}$, quark energy $E_f=\sqrt{p^2_3+2 n |Q_f B|+m_q^2}$, and Fermi-Dirac distribution function $F(x)=\left( e^{x/T}+1 \right)^{-1}$. At nonzero magnetic field, the three-dimensional quark momentum integration becomes a one-dimensional momentum integration and a summation over the discrete Landau levels.

In non-chiral limit, the chiral restoration at finite temperature is a smooth crossover, and the pseudo-critical temperature $T_{pc}$ is defined through the maximum decrease of (dynamical) quark mass, $\frac{\partial^2 m_q}{\partial T^2}=0$. As we know, with stronger coupling $G$ between quarks, the restoration of chiral symmetry happens at a higher pseudo-critical temperature $T_{pc}$.

In NJL model, mesons are constructed through quark bubble summations in the frame of random phase approximation~\cite{njl2,njl3,njl4,njl5,zhuang,maocharge},
\begin{equation}
{\cal D}_M(x,z)  = 2G \delta(x-z)+\int d^4y\ 2G \Pi_M(x,y) {\cal D}_M(y,z),
\label{dsequ}
\end{equation}
where ${\cal D}_M(x,z)$ represents the meson propagator from $x$ to $z$ in coordinate space, and the corresponding meson polarization function is the quark bubble,
\begin{equation}
\label{bubble}
\Pi_M(x,y) = i{\text {Tr}}\left[\Gamma_M^{\ast} S(x,y) \Gamma_M  S(y,x)\right]
\end{equation}
with the meson vertex
\begin{equation}
\label{vertex} \Gamma_M = \left\{\begin{array}{ll}
1 & M=\sigma\\
i\tau_+\gamma_5 & M=\pi_+ \\
i\tau_-\gamma_5 & M=\pi_- \\
i\tau_3\gamma_5 & M=\pi_0\ ,
\end{array}\right.
\Gamma_M^* = \left\{\begin{array}{ll}
1 & M=\sigma\\
i\tau_-\gamma_5 & M=\pi_+ \\
i\tau_+\gamma_5 & M=\pi_- \\
i\tau_3\gamma_5 & M=\pi_0\ ,
\end{array}\right. \nonumber
\end{equation}
the quark propagator matrix in flavor space $S=diag(S_u,\ S_d)$, and the trace in spin, color and flavor spaces.

According to the Goldstone's theorem, the pseudo-Goldstone mode of chiral symmetry breaking under external magnetic field is the neutral pion $\pi^0$~\cite{gold1,gold2}. The charged pions $\pi^\pm$ are no longer degenerate with neutral pion $\pi^0$ since their direct interaction with magnetic fields breaks the isospin symmetry.

\subsubsection{neutral pion $\pi^0$}
The neutral pion $\pi^0$ is affected by external magnetic field only through the charged constituent quarks, and its propagator in momentum space can be derived as~\cite{mao2,maocharge,maopion}
\begin{equation}
\label{npole}
{\cal D}_{\pi^0}(k)=\frac{2G}{1-2G\Pi_{\pi^0}(k)},
\end{equation}
with the conserved momentum $k=(k_0, {\bf k})$ and polarization function $\Pi_{\pi^0}(k)$ of $\pi^0$ meson under external magnetic field. Near the pole, we have
\begin{equation}
1-2G\Pi_{\pi^0}(k)=\left(k_0^2-E^2_{\pi^0}({\bf k}^2)\right)\times {\text {const}},
\end{equation}
and the $\pi^0$ meson energy is given as $E^2_{\pi^0}({\bf k}^2)=v^2_\parallel k^2_3+v^2_\perp (k^2_1+k^2_2) +m_{\pi^0}^2$, with longitudinal and transverse velocity $v_{\parallel, \perp}$ and pole mass $m_{\pi^0}$ of $\pi^0$ meson under external magnetic field.

Therefore, the $\pi^0$ meson pole mass $m_{\pi^0}$ is defined as the pole of the propagator at zero momentum ${\bf k}={\bf 0}$,
\begin{equation}
\label{mmass}
1-2G\Pi_{\pi^0}(k_0^2=m^2_{\pi^0}, {\bf k}^2=0)=0,
\end{equation}
and the polarization function can be simplified
\begin{eqnarray}
\label{pi}
\Pi_{\pi^0}(k_0^2,0) = J_1(m_q)+k_0^2 J_2(k_0^2)
\end{eqnarray}
with
\begin{eqnarray}
J_2(k_0^2) &=& 3\sum_{f,n}\alpha_n \frac{|Q_f B|}{2\pi} \int \frac{d p_3}{2\pi}{{1-2F(E_f)}\over  E_f (4 E_f^2-k_0^2)}.\nonumber
\end{eqnarray}

\subsubsection{charged pions $\pi^\pm$}
When constructing charged mesons through quark bubble summations, we should take into account of the interaction between charged mesons and magnetic fields. The charged pions $\pi^\pm$ with zero spin are Hermite conjugation to each other, and they have the same mass at finite temperature and magnetic field.

The $\pi^+$ meson propagator ${\cal D}_{\pi^+}$ can be expressed in terms of the polarization function $\Pi_{\pi^+}$~\cite{maocharge,maopion},
\begin{eqnarray}
\label{eq4}
{\cal D}_{\pi^+}({\bar k})=\frac{2G}{1-2G\Pi_{\pi^+}({\bar k})},
\end{eqnarray}
where ${\bar k} =(k_0,0,-\sqrt{(2l+1)eB},k_3)$ is the conserved Ritus momentum of $\pi^+$ meson under magnetic fields. Near the pole, we have
\begin{equation}
1-2G\Pi_{\pi^+}({\bar k})=\left(k_0^2-E^2_{\pi^+}({\bf {\bar k}}^2)\right)\times {\text {const}},
\end{equation}
and the $\pi^+$ meson energy is given as $E^2_{\pi^+}({\bf {\bar k}}^2)=v^2_\parallel k^2_3+(2l+1)|eB| +m_{\pi^+}^2$, with longitudinal velocity $v_{\parallel}$, Landau level $l$ and pole mass $m_{\pi^+}$ of $\pi^+$ meson under external magnetic field.

The $\pi^+$ meson pole mass $m_{\pi^+}$ is defined through the pole of the propagator at zero momentum $(l=0,\ k_3=0)$,
\begin{eqnarray}
1-2G\Pi_{\pi^+}(k_0^2=|eB|+m^2_{\pi^+})=0,
\label{pip}
\end{eqnarray}
and
\begin{eqnarray}
\Pi_{\pi^+}(k_0^2) &=& J_1(m_q)+J_3(k_0^2),\label{eq7}\\
J_3(k_0^2) &=& \sum_{n,n'} \int \frac{d p_3}{2\pi}\frac{j_{n,n'}(k_0^2)}{4E_n E_{n'}} \times \label{eq8}\\
&& \Big[\frac{F(-E_{n'})- F(E_n)}{k_0+E_{n'}+E_n}+\frac{F(E_{n'})- F(-E_n)}{k_0-E_{n'}-E_n}\nonumber\\ &&+\frac{F(-E_{n})- F(-E_{n'})}{k_0+E_{n'}-E_n}+\frac{F(E_{n})- F(E_{n'})}{k_0-E_{n'}+E_n}\Big],\nonumber\\
j_{n,n'}(k_0^2)& = &\left[{(k_0)^2/2}-n'|Q_u B|-n|Q_d B|\right]j^+_{n,n'} \nonumber\\
&&-2 \sqrt{n'|Q_u B|n|Q_d B|}\ j^-_{n,n'},\label{eq9}
\end{eqnarray}
with $u$-quark energy $E_{n'}=\sqrt{p^2_3+2 n' |Q_u B|+m_q^2}$, $d$-quark energy $E_n=\sqrt{p^2_3+2 n |Q_d B|+m_q^2}$, and summations over Landau levels of $u$ and $d$ quarks in $J_3(k_0^2)$.

\section{results and discussions}
\label{sec:r}

Because of the four-fermion interaction, the NJL model is not a renormalizable theory and needs regularization. In this work, we make use of the gauge invariant Pauli-Villars regularization scheme~\cite{njl1,njl2,njl3,njl4,njl5,mao,maocharge,mao2,mao1,mao11,maopion,limao2023}, where the quark momentum runs formally from zero to infinity. To consider the same physical situation as in LQCD simulations~\cite{ding200800493,ding2022}, the three parameters in our Pauli-Villars regularized NJL model, namely the current quark mass $m_0=14.6$ MeV, the coupling constant $G_0=3.62$ GeV$^{-2}$ and the Pauli-Villars mass parameter $\Lambda=1094$ MeV are fixed by fitting the chiral condensate $\langle\bar\psi\psi\rangle=-(250\ \text{MeV})^3$, and LQCD results~\cite{ding200800493,ding2022} of pion mass $m_\pi=220$ MeV and pion decay constant $f_\pi=96.9$ MeV at $T=0$ and $eB=0$. With these parameters in the NJL model, the pseudo-critical temperature of chiral restoration phase transition at vanishing magnetic field is $T_{pc}=169$ MeV, which is almost the same as the LQCD result $T_{pc}\simeq 170$ MeV~\cite{ding2022}.

In this paper, we determine $G(eB)=G^0(eB)$ $(G(eB)=G^+(eB))$ through the neutral (charged) pion mass spectra from LQCD simulations at vanishing temperature and finite magnetic field~\cite{ding200800493}, and calculate the pseudo-critical temperature $T_{pc}(eB)$ of chiral restoration phase transition under external magnetic field by using obtained $G(eB)$.

\subsection{$G^0(eB)$ fitted from the $\pi^0$ mass spectra}

\begin{figure}[hbt]
\centering
\includegraphics[width=7cm]{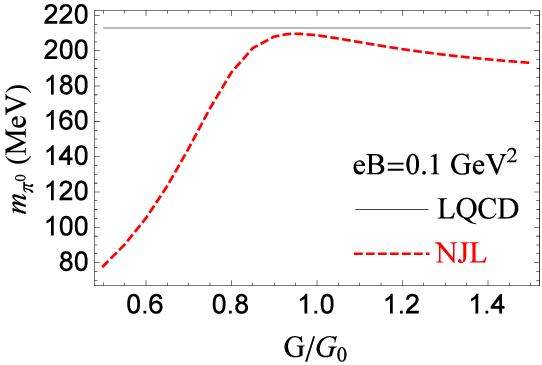}
\includegraphics[width=7cm]{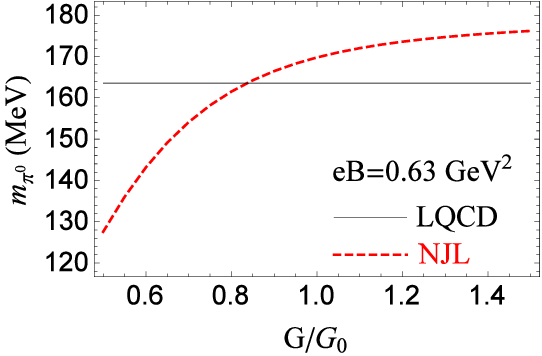}
\includegraphics[width=7cm]{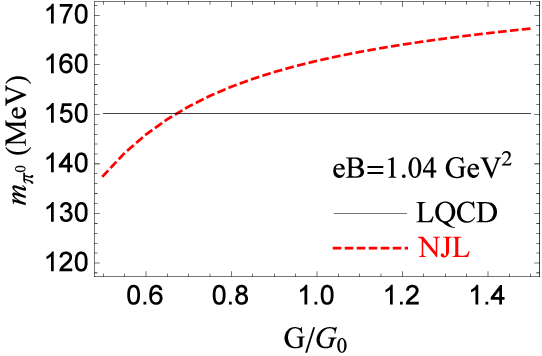}
\caption{$\pi^{0}$ meson mass $m_{\pi^0}$ as a function of coupling constant $G/G_0$ in the NJL model (in red) and the LQCD results of $m_{\pi^0}$ (in black)~\cite{ding200800493} with vanishing temperature $T=0$ and fixed magnetic field $eB=0.1,\ 0.63,\ 1.04$ GeV$^2$.} \label{fmpi0}
\end{figure}
\begin{figure}[hbt]
\centering
\includegraphics[width=7cm]{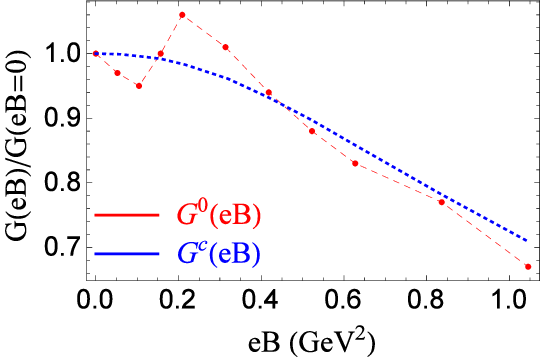}
\includegraphics[width=7cm]{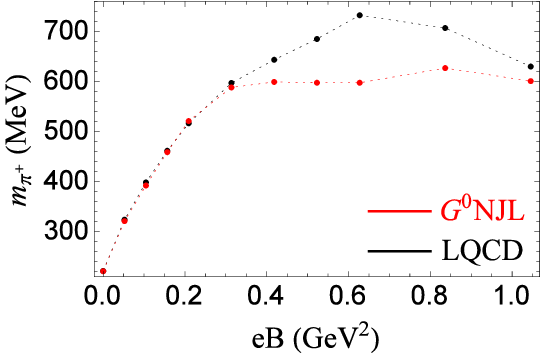}
\includegraphics[width=7cm]{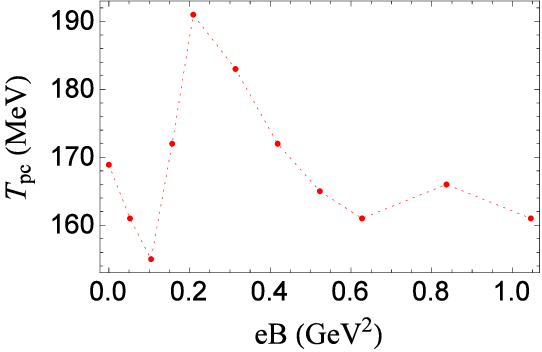}
\caption{(upper panel) Magnetic field dependent coupling constant $G^0(eB)$ in the NJL model fitted from LQCD results of $\pi^{0}$ meson mass $m_{\pi^0}$~\cite{ding200800493} with vanishing temperature and finite magnetic field in the red line and $G^c(eB)$ determined by the constant pseudo-critical temperature $T_{pc}(eB)=T_{pc}(eB=0)$ of chiral restoration phase transition in the blue line. (middle panel) The mass spectra of $\pi^+$ meson $m_{\pi^+}$ as a function of magnetic field at vanishing temperature calculated by using $G^0(eB)$ in the NJL model (in red) and the LQCD result of $m_{\pi^+}$ with vanishing temperature and finite magnetic field (in black)~\cite{ding200800493}. (lower panel) The pseudo-critical temperature $T_{pc}$ of chiral restoration phase transition as a function of magnetic field calculated by using $G^0(eB)$ in the NJL model. } \label{fgebpi0}
\end{figure}

Figure~\ref{fmpi0} plots the $\pi^{0}$ meson mass $m_{\pi^0}$ as a function of coupling constant $G/G_0$ calculated in the NJL model (in red) and the LQCD results of $m_{\pi^0}$ (in black)~\cite{ding200800493} with vanishing temperature $T=0$ and fixed magnetic field $eB=0.1,\ 0.63,\ 1.04$ GeV$^2$. At weak magnetic fields, such as $eB=0.1$ GeV$^2$, $m_{\pi^0}$ shows non-monotonic behavior with the change of coupling between quarks. As $G$ decreases, it firstly increases and then decreases. It is noticeable that the LQCD result can not be approached exactly. In this case, we fix $G(eB)=0.95 G_0$ by the peak value of $m_{\pi^0}$.  At stronger magnetic fields, such as $eB=0.63,\ 1.04$ GeV$^2$, $\pi^{0}$ mass decreases with decreasing coupling constant $G$, and it meets the LQCD result at $G(eB)=0.83 G_0$ and $G(eB)=0.67 G_0$, respectively.

The magnetic field dependent coupling constant $G^0(eB)$ constrained by the $\pi^{0}$ meson mass of LQCD simulations with vanishing temperature and finite magnetic field~\cite{ding200800493} is shown in the red line of Fig.\ref{fgebpi0} upper panel, which firstly decreases, then increases and finally decreases as magnetic fields grow. The mass spectra of $\pi^+$ meson $m_{\pi^+}$ as a function of magnetic field at vanishing temperature is calculated by using magnetic field dependent coupling constant $G^0(eB)$ in the NJL model, see the red line of Fig.\ref{fgebpi0} middle panel. The LQCD results of $m_{\pi^+}$~\cite{ding200800493} are plotted in the black line of Fig.\ref{fgebpi0} middle panel. With $eB< 0.3$ GeV$^2$, $m_{\pi^+}$ increases monotonically with magnetic fields, which is consistent with LQCD results. With $0.4$ GeV$^2$ $\leq eB \leq 0.8$ GeV$^2$, $m_{\pi^+}$ slightly decreases with magnetic fields and then increases, which deviates from LQCD results. With $0.8$ GeV$^2$ $\leq eB\leq 1.05$ GeV$^2$, $m_{\pi^+}$ decreases with magnetic fields, which is quantitatively deviated from LQCD results, but qualitatively consistent with LQCD results. For reference, we depict in blue a $G^c(eB)$ in Fig.\ref{fgebpi0} upper panel, which solves a constant pseudo-critical temperature for chiral restoration phase transition under external magnetic field $T_{pc}(eB)=T_{pc}(eB=0)$. As we know, the pseudo-critical temperature is enhanced by the external magnetic field, due to the dimension reduction of quarks, and it increases with increasing coupling between quarks. Comparing $G^0(eB)$ with $G^c(eB)$, it is expected that the $G^0(eB)$ in NJL model will lead to a non-monotonic behavior of the pseudo-critical temperature $T_{pc}(eB)$ of chiral restoration phase transition, as shown explicitly in Fig.\ref{fgebpi0} lower panel. This is different from the LQCD result of a decreasing $T_{pc}(eB)$~\cite{ding2022}.

Let's have a short summary for this part. We determine a magnetic field dependent coupling constant $G^0(eB)$ in the NJL model by the LQCD results of $\pi^{0}$ meson mass at vanishing temperature and finite magnetic field~\cite{ding200800493}, which shows non-monotonic behavior at weak magnetic field region, and decreases with magnetic fields at strong magnetic field region. By using such $G^0(eB)$ in the NJL model, we calculate $\pi^{+}$ meson mass $m_{\pi^+}(eB)$ at vanishing temperature and finite magnetic field and the pseudo-critical temperature $T_{pc}(eB)$ of chiral restoration phase transition under external magnetic field. $m_{\pi^+}(eB)$ reproduces the LQCD results~\cite{ding200800493} in weak magnetic field region, but shows deviations in strong magnetic field region. $T_{pc}(eB)$ is a non-monotonic function of magnetic fields, which is different from the LQCD result~\cite{ding2022}.

\subsection{$G^+(eB)$ fitted from the $\pi^+$ mass spectra}

\begin{figure}[hbt]
\centering
\includegraphics[width=7cm]{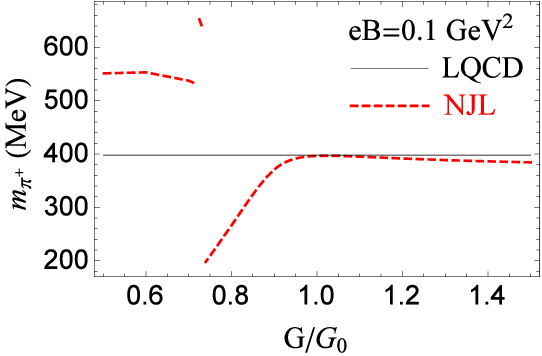}
\includegraphics[width=7cm]{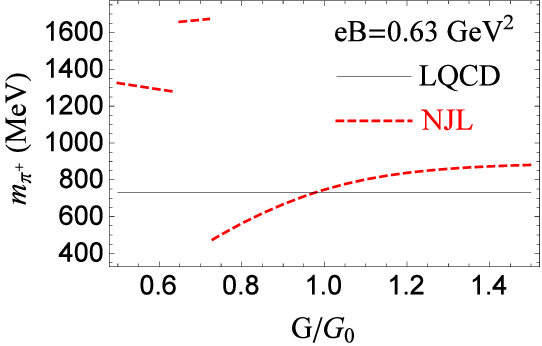}
\includegraphics[width=7cm]{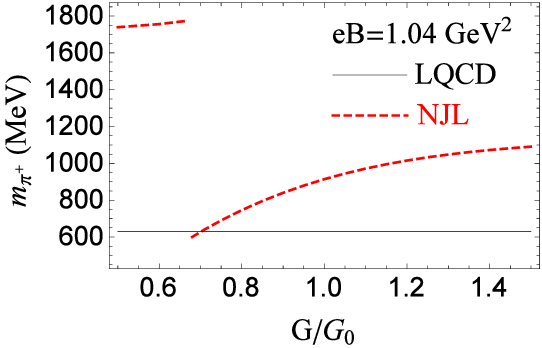}
\caption{$\pi^{+}$ meson mass $m_{\pi^+}$ as a function of coupling constant $G/G_0$ in the NJL model (in red) and the LQCD results of $m_{\pi^+}$ (in black)~\cite{ding200800493} with vanishing temperature $T=0$ and fixed magnetic field $eB=0.1,\ 0.63,\ 1.04$ GeV$^2$.} \label{fmpicharge}
\end{figure}
\begin{figure}[hbt]
\centering
\includegraphics[width=7cm]{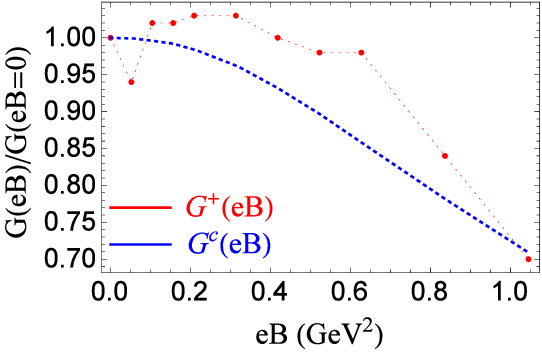}
\includegraphics[width=7cm]{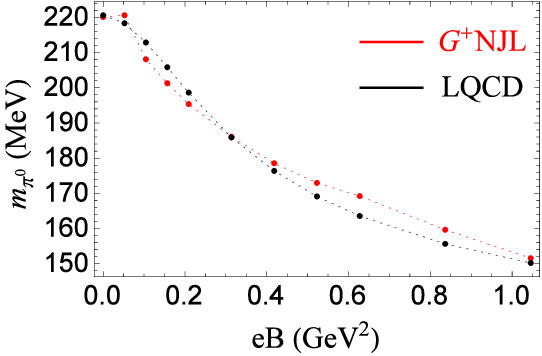}
\includegraphics[width=7cm]{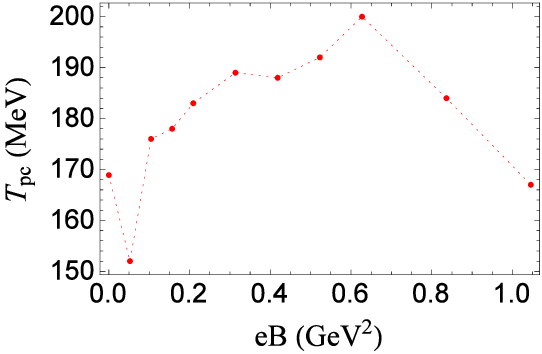}
\caption{(upper panel) Magnetic field dependent coupling constant $G^+(eB)$ in the NJL model fitted from LQCD results of $\pi^{+}$ meson mass spectra $m_{\pi^+}$~\cite{ding200800493} in the red line and $G^c(eB)$ determined by the constant pseudo-critical temperature $T_{pc}(eB)=T_{pc}(eB=0)$ of chiral restoration phase transition in the blue line. (middle panel) The mass spectra of $\pi^0$ meson as a function of magnetic field at vanishing temperature calculated by using $G^+(eB)$ in the NJL model (in red) and the LQCD result of $m_{\pi^0}$ (in black)~\cite{ding200800493}. (lower panel) The pseudo-critical temperature $T_{pc}$ of chiral restoration phase transition as a function of magnetic field calculated by using $G^+(eB)$ in the NJL model.} \label{fgebpicharge}
\end{figure}
Figure~\ref{fmpicharge} depicts $\pi^{+}$ meson mass $m_{\pi^+}$ as a function of coupling constant $G/G_0$ calculated in the NJL model (in red) and the LQCD results of $m_{\pi^+}$ (in black)~\cite{ding200800493} with vanishing temperature $T=0$ and fixed magnetic field $eB=0.1,\ 0.63,\ 1.04$ GeV$^2$. At weak magnetic fields, such as $eB=0.1$ GeV$^2$, $m_{\pi^+}$ shows non-monotonic behavior with the change of coupling between quarks. As $G$ decreases, it firstly increases and then decreases. At weak coupling region, we observe the mass jumps of $\pi^{+}$ meson, caused by the discrete energy level of constituent quarks~\cite{maocharge}. It is noticeable that the LQCD result can not be approached exactly, and we have to define the $G(eB)=1.02 G_0$ by the peak value of $m_{\pi^+}$ in this case. At stronger magnetic fields $eB=0.63,\ 1.04$ GeV$^2$, $m_{\pi^+}$ decreases with decreasing $G$, which crosses the LQCD values at $G(eB)=0.98 G_0$ and $G(eB)=0.7 G_0$, respectively. In these cases, we also observe the mass jumps of $\pi^{+}$ meson at weak coupling region, caused by the discrete energy level of constituent quarks~\cite{maocharge}.

The resulting magnetic field dependent coupling constant $G^+(eB)$ fitted from LQCD results of $m_{\pi^+}$ with vanishing temperature and finite magnetic field~\cite{ding200800493} is depicted in the red line of Fig.\ref{fgebpicharge} upper panel, which shows non-monotonic behavior at weak magnetic field region, and decreases with magnetic fields at strong magnetic field region. In the middle panel of Fig.\ref{fgebpicharge}, we calculate the $\pi^{0}$ meson mass $m_{\pi^0}$ as a function of magnetic field at vanishing temperature by using $G^+(eB)$ in the NJL model. $m_{\pi^0}$ slightly increases at very weak magnetic field and then decreases at stronger magnetic field. Comparing with the LQCD results (in black), except for the very weak magnetic field region, our results are reasonably good. For reference, $G^c(eB)$ solving a constant pseudo-critical temperature $T_{pc}(eB)=T_{pc}(eB=0)$ of chiral restoration phase transition is also plotted in Fig.\ref{fgebpicharge} upper panel. It is indicated that the $G^+(eB)$ in NJL model will lead to a non-monotonic behavior for the pseudo-critical temperature $T_{pc}(eB)$ of chiral restoration phase transition under external magnetic field. As shown explicitly in the lower panel of Fig.\ref{fgebpicharge}, $T_{pc}(eB)$ mainly increases with magnetic fields associated with some oscillations at $eB<0.6$ GeV$^2$, which is different from the LQCD result~\cite{ding2022}, and $T_{pc}(eB)$ decreases with magnetic fields at $eB>0.6$ GeV$^2$, which is consistent with LQCD conclusion~\cite{ding2022}.

Parallel with the last section, we constrain a magnetic field dependent coupling constant $G^+(eB)$ in the NJL model by the LQCD results of $\pi^{+}$ meson mass at vanishing temperature and finite magnetic field~\cite{ding200800493}. $G^+(eB)$ shows non-monotonic behavior at weak magnetic field region, and decreases with magnetic fields at strong magnetic field region. By using such $G^+(eB)$ in the NJL model, we calculate $\pi^{0}$ meson mass $m_{\pi^0}(eB)$ at vanishing temperature and finite magnetic field and the pseudo-critical temperature $T_{pc}(eB)$ of chiral restoration phase transition under external magnetic field. $m_{\pi^0}(eB)$ reproduces the LQCD results quite well except for the region with very weak magnetic field. $T_{pc}(eB)$ is a non-monotonic function of magnetic fields. The decreasing $T_{pc}(eB)$ at strong magnetic field region is consistent with LQCD results~\cite{ding2022}.

\section{summary}
\label{sec:s}
The magnetic field dependent coupling constant $G(eB)$ is considered in a two-flavor NJL model. Based on LQCD results of the pion mass spectra under external magnetic field, we constrain the $G(eB)$, and calculate the corresponding pseudo-critical temperature of chiral restoration phase transition with obtained $G(eB)$.

On one side, the magnetic field dependent coupling constant $G^0(eB)$ is determined by the LQCD results of neutral pion mass spectra with vanishing temperature and finite magnetic field, and it is a non-monotonic function of magnetic field. By using such $G^0(eB)$ in the NJL model, we find that it reproduces the LQCD results of charged pion mass spectra in weak magnetic field region, but leads to a non-monotonic behavior of the pseudo-critical temperature of chiral restoration phase transition under external magnetic field.

On the other side, the magnetic field dependent coupling constant $G^+(eB)$ is determined by the LQCD results of charged pion mass spectra with vanishing temperature and finite magnetic field, and it is a non-monotonic function of magnetic field, too. By using such $G^+(eB)$ in the NJL model, we find that it reproduces the LQCD results of neutral pion mass spectra except for very weak magnetic field region, but leads to a non-monotonic behavior of the pseudo-critical temperature of chiral restoration phase transition under external magnetic field. Note that $G^0(eB)$ and $G^+(eB)$ are different from each other.

In previous work~\cite{geb1,meson,geb3,geb4,bf8,bf9,su3meson4,mao2,limao2023}, the magnetic field dependent coupling constant $G(eB)$ has been fitted by the LQCD results of order parameter or pseudo-critical temperature of chiral restoration phase transition, which leads to a decreasing coupling constant under external magnetic field. With such $G(eB)$ in the NJL model, the mass spectra of neutral and charged pions can be qualitatively described in weak magnetic field region, but the saturation behavior of neutral pion mass spectra and decreasing behavior of charged pion mass spectra under strong magnetic field can not be explained.

In conclusion, although the coupling between quarks plays an important role in determining the chiral symmetry breaking/restoration and light meson mass spectra in the NJL model, a magnetic field dependent coupling constant $G(eB)$ can not simultaneously explain the reduction of pseudo-critical temperature of chiral restoration phase transition and the light meson mass spectra under external magnetic field. To comprehensively understand the QCD phase structure under external magnetic field, other physical factors are needed to be introduced.\\

\noindent {\bf Acknowledgement:}
Many thanks go to Prof. H.T. Ding in CCNU, who kindly provides us their LQCD data~\cite{ding200800493} of the pion mass spectra under external magnetic field. The work is supported by the NSFC grant 12275204.

\end{document}